# Dual-Mode Luminescent Thermometry in LiYO$_2$:Nd$^{3+}$,Yb$^{3+}$ Enabled by Structural Phase Transition and Phonon-Assisted Energy Transfer

M. Tahir Abbas[1], M. Szymczak[1], D. Szymanski[1], M. Drozd[1], G. Chen[2], L. Marciniak[1*]

[1] Institute of Low Temperature and Structure Research, Polish Academy of Sciences, Okólna 2, 50-422 Wrocław, Poland

[2] Key Laboratory of Micro-systems and Micro-structures, Ministry of Education, Harbin Institute of Technology, Harbin, 150001, China

*corresponding author: l.marciniak@intibs.pl

*KEYWORDS: luminescence thermometry, phonon-assisted energy transfer, ratiometric approach, luminescence kinetics, phase transition*

**Abstract**

In this work, a dual-mode luminescent thermometer operating via both ratiometric and lifetime-based readout strategies was developed, enabled by the coexistence of two thermally driven effects: a structural phase transition in LiYO$_2$ and a phonon-assisted energy transfer from Yb$^{3+}$ to Nd$^{3+}$. As demonstrated, changes in the shape of the emission band of Yb$^{3+}$ ions corresponding to the $^2F_{5/2} \rightarrow {}^2F_{7/2}$ electronic transition, induced by the phase transition, allowed the design of a ratiometric thermometer with a maximum relative sensitivity ($S_R$) of 3.1% K$^{-1}$ for LiYO$_2$:10%Yb$^{3+}$, 1% Nd$^{3+}$ at 290 K. On the other hand, the temperature-dependent Yb$^{3+} \rightarrow$ Nd$^{3+}$ energy transfer facilitated the development of a lifetime-based thermometer with a maximum $S_R$ of 1.5% K$^{-1}$ for 20% Nd$^{3+}$ at 378 K. In both approaches, tuning the Nd$^{3+}$ ions concentration enabled modulation of both the sensitivity and the temperature at which the maximum $S_R$ was achieved. This was accomplished by shifting the phase transition temperature and enhancing the probability of interionic energy transfer, respectively. Notably, the temperature ranges corresponding to the maximum $S_R$ for the ratiometric and lifetime modes were found to be distinct, effectively broadening the thermal operating range of the sensor. Additionally, it was shown that LiYO$_2$:Nd$^{3+}$, Yb$^{3+}$ can be also used as a temperature sensor through the ratio of luminescence intensities recorded at two distinct time gates. Furthermore, the results confirmed that the phonon-assisted energy transfer process plays a dominant role in shaping the luminescence kinetics, surpassing the influence of the structural phase transition. Overall, this study highlights LiYO$_2$:Nd$^{3+}$,Yb$^{3+}$ as a promising candidate for multimodal luminescent temperature sensing applications.

**Introduction**

Luminescence thermometry has gained significant recognition in recent years due to its unique advantages, including rapid response, remote measurements, and reliable performance in harsh or extreme environments that are subject to electromagnetic interference and strong corrosion[1–6]. These unique features mitigate the inherent limitations of conventional contact-based sensing methods, facilitating precise and reliable remote thermal mapping and underscoring their strong potential for applications in multiple disciplines[2,7]. Among various strategies that can be used to determine the temperature in luminescence thermometry those based on luminescence intensity ratio (*LIR*) and luminescence kinetics constitute the most widely utilized owing to their high reliability[2,7]. Although this strategy is highly promising, the application of the ratiometric approach is constrained under certain conditions. In particular, when the phosphor is placed in a medium with strong light absorption or scattering, the dispersion can substantially distort the emission spectrum, thereby affecting the accuracy of the temperature readout[7]. To address this limitation either the ratiometric thermometer with small spectral separation between emission bands exploited to temperature sensing should be considered or kinetic based approach should be utilized.

Ratiometric thermometers designed with closely spaced emission bands often rely on thermally coupled energy levels or Stark components of excited states[8–10]. The relative sensitivity of such thermometers is proportional to the energy gap between these levels, which must remain below 2000 cm$^{-1}$ to preserve efficient thermal coupling. As a result, the relative sensitivity of these thermometers typically does not exceed 2% K$^{-1}$. When utilizing *LIR* between Stark components, where the energy separation is typically around 100 cm$^{-1}$, the sensitivity is even further reduced[11–13]. This limitation has led to the exploration of alternative strategies aimed at achieving higher relative sensitivity.

Among these, luminescence thermometry based on thermally induced phase transitions has garnered significant interest due to its superior thermometric performance[14–25]. Structural phase transitions alter the point symmetry of the crystallographic sites occupied by dopant ions, modifying the number and energy of Stark components and thereby reshaping the emission spectrum. The opposite thermal behavior of emission lines corresponding to low- and high-temperature phases of the phosphor enables the development of ratiometric luminescent thermometers with exceptionally high relative sensitivities, reaching values up to 35.2% K$^{-1}$ [24].

Alternatively, lifetime-based luminescence thermometry represents an attractive approach due to its insensitivity to changes in the extinction coefficient of the surrounding medium[2,7]. While this method shows strong potential, its relative sensitivity is generally lower than that of ratiometric techniques. Enhancing the performance of lifetime-based thermometers thus requires leveraging other thermally dependent energy transfer mechanisms. Recent studies indicate that phonon-assisted energy transfer offers a promising pathway to improve sensitivity[26–30]; however, the interplay between thermally induced structural phase transitions and phonon-assisted energy transfer remains largely unexplored.

To comprehensively address these considerations, this study investigates the spectroscopic properties of $LiYO_2:Nd^{3+},Yb^{3+}$ phosphors as a function of temperature, evaluating their potential for both ratiometric and lifetime-based luminescence thermometry. The co-doping of $Nd^{3+}$ and $Yb^{3+}$ ions enables a synergistic effect between phonon-assisted energy transfer and thermally induced phase transition. The impact of $Nd^{3+}$ ions concentration on thermometric performance is thoroughly examined. In addition, two distinct strategies are applied to analyze the luminescence kinetics of $Yb^{3+}$ ions: time-resolved luminescence decay profiling and a time-gated detection approach.

**Experimental Section**

*Materials*

$LiYO_2:10\%Yb^{3+}$, x% $Nd^{3+}$ (x=1, 5, 15, 20) phosphors were synthesized by using the conventional high-temperature solid state reaction method. $Li_2CO_3$ (99.9% of purity, Chempur), $Y_2O_3$ (99.999% of purity, Stanford Materials Corporation), $Nd_2O_3$ (99.99% of purity, Stanford Materials Corporation), and $Yb_2O_3$ (99.99% of purity, Stanford Materials Corporation) were used as starting materials. The samples were ground in an agate mortar with a few drops of hexane and then annealed in alumina crucibles at 1273 K for 6 hours with a heating rate of 10 K min$^{-1}$. After cooling to room temperature, the powder samples were ground again for subsequent characterization.

*Methods*

The obtained materials were examined by using the powder X-ray diffraction technique. Powder diffraction data were acquired in Bragg-Brentano geometry with a PANalytical X'Pert Pro diffractometer equipped with Oxford Cryosystems Phenix low-temperature and Anton Paar

HTK 1200 N high-temperature attachments, employing Ni-filtered Cu-Kα radiations (40 kV, 30 mA). Diffraction patterns were measured in 10 – 90º 2$\theta$ range. ICSD database entries 50992 (Low-temperature phase -LT) and 50993 (High-temperature phase - HT) were used as initial models for the analysis of the obtained diffraction data.

Scanning electron microscopy (SEM) was applied to evaluate the morphology of the samples, while elemental mapping was performed by energy-dispersive X-ray spectroscopy (EDS). The analyses were performed using an FEI Nova NanoSEM 230 instrument equipped with an EDAX Genesis XM4 spectrometer, with accelerating voltages of 5 and 30 kV for SEM and EDS mapping, respectively. The EDS map of the Li element was not recorded and discussed in this paper due to the limitations of the EDS method regarding the detection of light elements ($Z \leq 4$). For preparation, powdered samples were mounted onto a piece of conductive carbon tape, which was adhered to a metal stub. This procedure ensures stable positioning of the sample during SEM imaging while minimizing charging effects during electron beam exposure.

Differential scanning calorimetry (DSC) measurements were performed on a Perkin-Elmer DSC 8000 calorimeter equipped with a Controlled Liquid Nitrogen Accessory (LN2). Powdered samples were sealed in aluminum pans and measurements were conducted over a temperature range of 200-500 K at a heating and cooling rate of 20 K min$^{-1}$.

The excitation spectra were measured by using the FLS1000 Fluorescence Spectrometer from Edinburgh Instruments equipped with a 450 W Xenon lamp and R5509-72 photomultiplier tube from Hamamatsu in a nitrogen-flow cooled housing for near infrared range detection. The emission spectra were recorded under 808 nm laser diode excitation using the same system. To conduct the temperature dependent measurements, the temperature of the sample was controlled using a THMS600 heating-cooling stage from Linkam (0.1 K temperature stability and 0.1 K point resolution). Luminescence decay profiles were also measured with the same spectrometer, using a 150 W $\mu$Flash lamp. The measured luminescence decay profiles of $Yb^{3+}$ ions were fitted using double exponential function:

$$I(t) = I_0 + A_1 \cdot \exp\left(-\frac{t}{\tau_1}\right) + A_2 \cdot \exp\left(-\frac{t}{\tau_2}\right) \quad (1)$$

where $\tau_1$ and $\tau_1$ are the decay components and $A_1$ and $A_2$ are the amplitudes of the double-exponential function. Based on the obtained results the average lifetime ($\tau_{avr}$) of the excited state was calculated as follows:

$$\tau_{avr} = \frac{A_1\tau_1^2 + A_2\tau_2^2}{A_1\tau_1 + A_2\tau_2} \quad (2)$$

**Results and discussion**

$LiYO_2$ has long attracted attention due to its ability to adopt different crystal structures depending on external conditions. At ambient temperature, it can exist either in a monoclinic ($P2_1/c$ space group) or tetragonal ($I4_1/amd$ space group) form (Figure 1a), and the stabilized phase is strongly influenced by factors such as crystallite size or dopant ion and its concentration[31–35]. Temperature variations can further alter its structural arrangement. When the $LiYO_2$ is heated to around 370 K (single crystal), a reversible first-order phase transition takes place, giving rise to the tetragonal phase, which exhibits a markedly higher symmetry than the monoclinic one. This transformation involves both a change in the unit cell parameters and a modification of the $Y^{3+}$ local symmetry, shifting from $C_2$ to $D_{2d}$ point symmetry[31–35]. These structural changes have a profound impact on the optical properties of doped $LiYO_2$-based phosphors. In this case, $Yb^{3+}$ and $Nd^{3+}$ ions were introduced into the $LiYO_2$, preferentially occupying $Y^{3+}$ crystallographic positions due to their comparable ionic radii and equivalent charge states, which ensures minimal lattice distortion upon substitution. The comparison of room temperature XRD patterns of the $LiYO_2$:10%$Yb^{3+}$, $Nd^{3+}$ with different concentration of $Nd^{3+}$ ions reveals that for the $LiYO_2$:10%$Yb^{3+}$, 1%$Nd^{3+}$ the presence of the reflections assigned to both low temperature and high temperature phase of $LiYO_2$ can be found (Figure 1b). However, for higher $Nd^{3+}$ ions concentration only the low temperature phase can be found in the patterns. An explanation of this effect provide the DSC studies (Figure 1c), which revealed that for $LiYO_2$:10%$Yb^{3+}$, 1%$Nd^{3+}$ the phase transition can be found around 288 K, while an increase in $Nd^{3+}$ ions concentration results in a monotonic increase of the phase transition temperature ($T_{PT}$) up to 452 K for 20%$Nd^{3+}$ (Figure 1d). This effect associated with the strain induced in the host material structure associated with the difference in the ionic radii between host material cations and dopant ions was previously reported for many host materials[36–39]. The morphological studies of $LiYO_2$:$Nd^{3+}$,$Yb^{3+}$ revealed that the synthesized phosphors consists of particles of the average size around 7 μm (Figure 1e) and are characterized by homogenous distribution of ions (Figure 1f-i).

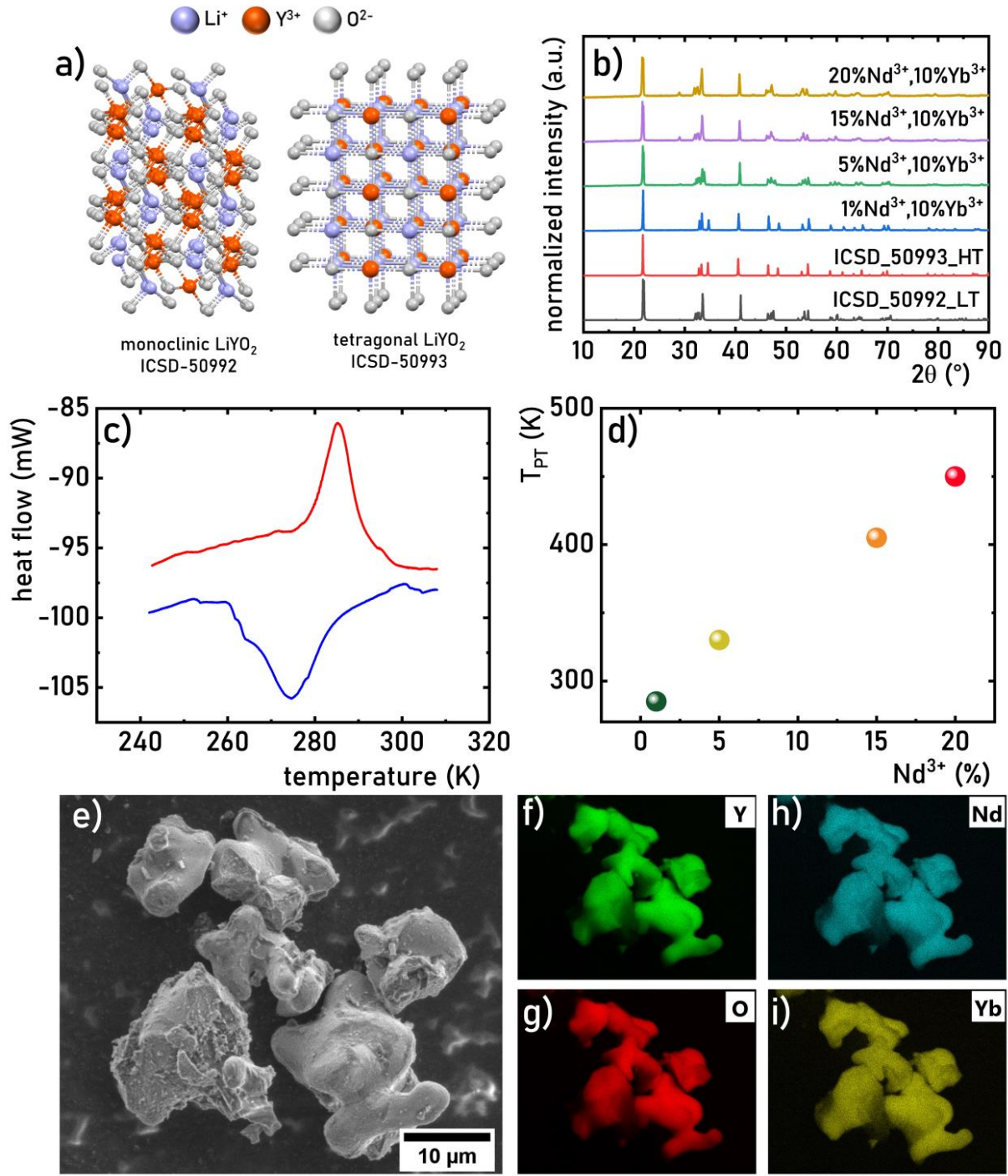

**Figure 1.** Structures of monoclinic and tetragonal structures of LiYO$_2$ – a); comparison of room temperature XRD patterns of LiYO$_2$:10%Yb$^{3+}$, Nd$^{3+}$ with different Nd$^{3+}$ concentrations – b); DSC curves of LiYO$_2$:10%Yb$^{3+}$, 1%Nd$^{3+}$ - c); the influence of Nd$^{3+}$ dopant concentration on $T_{PT}$ – d); the representative SEM images of LiYO$_2$:10%Yb$^{3+}$, 15%Nd$^{3+}$ - e); corresponding elemental distribution of Y – f), O – g), Nd – h) and Yb – i).

The luminescence behavior of LiYO$_2$:Nd$^{3+}$,Yb$^{3+}$ can be discussed using simplified energy level diagrams of Nd$^{3+}$ and Yb$^{3+}$ ions, as illustrated in Figure 2a. Under the laser excitation of $\lambda_{exc}$=808 nm the electrons are transferred from the $^4I_{9/2}$ ground state of Nd$^{3+}$ ions to the $^4F_{5/2}$ excited level, followed by the nonradiative depopulation to the $^4F_{3/2}$ metastable state. The radiative depopulation of the $^4F_{3/2}$ state leads to the occurrence of the emission at 1065 nm corresponding to the $^4F_{3/2} \rightarrow {}^4I_{11/2}$ electronic transition (except that the $^4F_{3/2} \rightarrow {}^4I_{9/2}$ and $^4F_{3/2} \rightarrow {}^4I_{13/2}$ electronic transition lead to the generation of the emission bands at ~880 nm and 1350 nm, respectively, which are not analyzed in this work). However, due to the small energy difference between $^4F_{3/2}$ state of Nd$^{3+}$ ions and $^2F_{5/2}$ excited state of Yb$^{3+}$ ions (~ 800 cm$^{-1}$) the phonon assisted energy transfer from $^4F_{3/2}$ enables the population of the $^2F_{5/2}$ state[40]. Consequently, the radiative depopulation of the $^2F_{5/2}$ leads to the generation of the emission band of Yb$^{3+}$ ions corresponding to the $^2F_{5/2} \rightarrow {}^2F_{7/2}$ electronic transition. Due to the large energy separation between $^2F_{5/2}$ and $^2F_{7/2}$ states of Yb$^{3+}$ ions (~10,000 cm$^{-1}$) the $^2F_{5/2}$ is not sensitive to its multiphonon depopulation and hence, Yb$^{3+}$ luminescence is considered as a thermally stable[41–45]. This effect was verified and confirmed for different host materials by the invaluable work of professor Georges Boulon [46–58]. The Nd$^{3+}\rightarrow$Yb$^{3+}$ energy transfer occurs with emission of phonons which makes it more probable than processes with phonon absorption. However, at elevated temperature the probability of such a Yb$^{3+}\rightarrow$Nd$^{3+}$ back energy transfer ($W_{BET}$) increases[13,40]. It results in both decrease of the Yb$^{3+}$ emission intensity and shortening of the lifetime of the $^2F_{5/2}$ state. In the case of the LiYO$_2$:Nd$^{3+}$,Yb$^{3+}$ the $\lambda_{exc}$=808 nm excitation wavelength results in the occurrence of the intense emission band at 980 nm corresponding to the $^2F_{5/2}\rightarrow{}^2F_{7/2}$ electronic transition of Yb$^{3+}$ ions and less intense emission band of Nd$^{3+}$ ions at 1060 nm corresponding to the $^4F_{3/2}\rightarrow{}^4I_{11/2}$ electronic transition (Figure 2b). At low temperatures (83 K), the emission spectra exhibit well-defined, sharp lines corresponding to electronic transitions between the Stark sublevels of the excited and ground states of both $Yb^{3+}$ and $Nd^{3+}$ ions. As the temperature increases above 403 K, these emission lines not only become spectrally broadened, which is a typical thermal effect, but also undergo shifts in their spectral positions. This behavior is attributed to changes in the local crystal field symmetry around the $Yb^{3+}$ and $Nd^{3+}$ ions, induced by a thermally driven structural phase transition. The excitation spectra recorded at an emission wavelength of 1060 nm (corresponding to the $Nd^{3+}$ ions luminescence) reveal numerous absorption bands associated with electronic transitions from the ground $^4I_{9/2}$ level of $Nd^{3+}$ ions to higher-energy excited states ($^4D_{3/2,11/2}$ (360 nm), $^2P_{11/2}$ (435 nm), $^2G_{9/2}$, $^4G_{11/2}$ (472 nm), $^4G_{9/2}$ (518 nm), $^4G_{7/2}$ + $^2K_{13/2}$ ,(532 nm,544 nm), $^4G_{5/2}$ + $^2G_{7/2}$ (580 nm) , $^4F_{7/2}$ (748 nm, 752 nm), $^4F_{5/2}$ (818 nm)) (Figure 2c). Similar to the emission spectra, the spectral positions of individual absorption lines in the excitation spectra shift following the phase transition, further indicating a modification of the local environment of the dopant ions. In the excitation spectra of $Yb^{3+}$ ions (recorded at $\lambda_{em}$=970 nm), similar absorption bands are observed as in the excitation spectra recorded at emission wavelength $\lambda_{em}$=1060 nm. This spectral similarity provides direct evidence of efficient energy transfer from $Nd^{3+}$ to $Yb^{3+}$ ions. Since the probability of the $W_{ET}$ $Nd^{3+} \rightarrow Yb^{3+}$ and $W_{BET}$ $Yb^{3+} \rightarrow Nd^{3+}$ depends on the interionic distance between interacting ions, the luminescent properties of $LiYO_2$:$Nd^{3+}$,$Yb^{3+}$ were investigated as a function of dopant concentration (Figure 2d). Additionally, as it was already discussed the higher amount of $Nd^{3+}$ ions leads to the increase the phase transition temperature. The comparison of the emission spectra of $LiYO_2$:$Nd^{3+}$,$Yb^{3+}$ with different amount of $Nd^{3+}$ ions measured at 83 K reveals that with increase of the $Nd^{3+}$ ions concentration the contribution of the luminescence lines associated with the HT phase increases. This is especially clear when the detailed analysis of the intensity of particular lines is performed. In the short wavelength part of the spectrum for low dopant concentration the emission line at 970 nm is dominant. However, when the $Nd^{3+}$ ions increases above 15% the additional line at 974 nm starts to be dominant. Additionally, the intensity of this line increases significantly with the line at 998 nm. This is an additional evidence that the contribution of the HT phase increases in the LiYO$_2$:Nd$^{3+}$,Yb$^{3+}$ at low temperature. The analysis of the kinetics of Yb$^{3+}$ ions emission revealed that with an increase in the Nd$^{3+}$ ions concentration the luminescence decay profile measured at $\lambda_{em}$=972 nm undergoes shortening (Figure 2e). The average lifetime of the $^2F_{5/2}$ state shortens from 0.52 ms for 1% Nd$^{3+}$ to 0.030 ms for 20% Nd$^{3+}$ as shown in Figure 2f. This shortening of $\tau_{avr}$ reflects efficient phonon-assisted Yb$^{3+}$ → Nd$^{3+}$ energy transfer, enabled by reduction of average interionic distances as the Nd$^{3+}$ ions concentration increases.

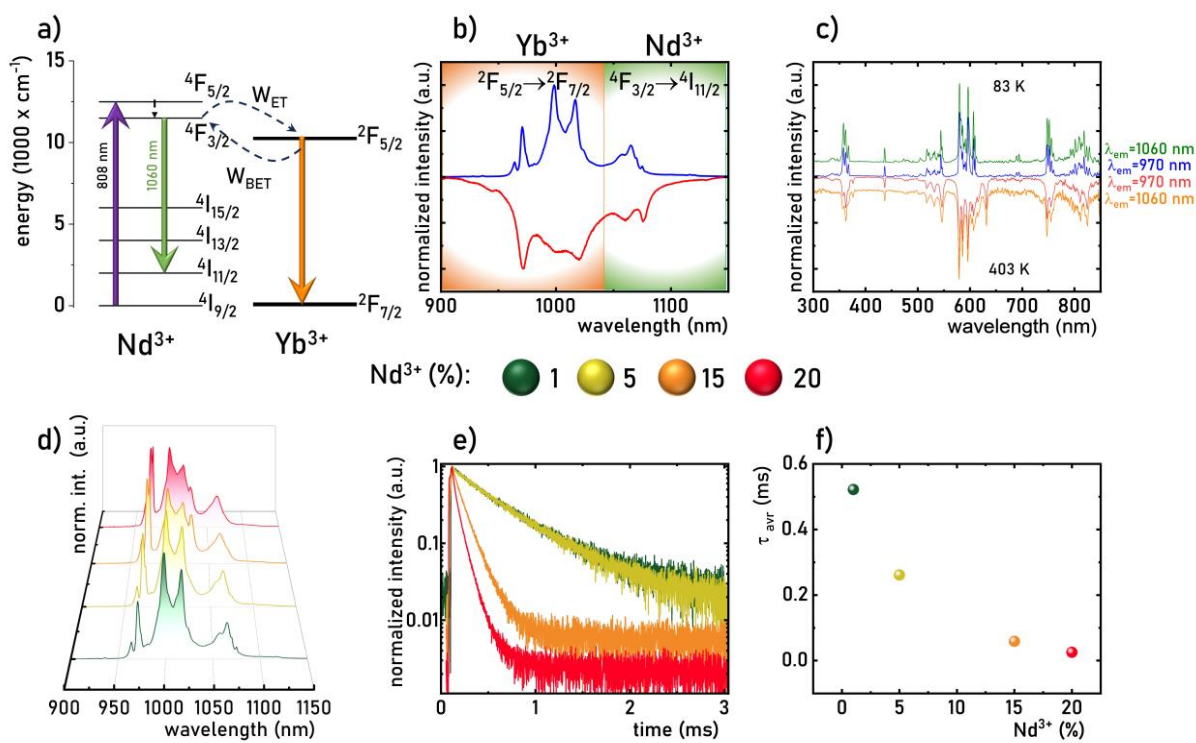

**Figure 2.** Simplified energy level diagram of Nd$^{3+}$,Yb$^{3+}$ ions - a); comparison of emission spectra of LiYO$_2$:1%Nd$^{3+}$,10%Yb$^{3+}$ measured at 83 K and 403 K corresponding to LT and HT phases of LiYO$_2$ - b); comparison of excitation spectra of LiYO$_2$:1%Nd$^{3+}$,10%Yb$^{3+}$ measured at 83 K and 403K for $\lambda_{em}$=970 nm (Yb$^{3+}$) and $\lambda_{em}$=1060 nm (Nd$^{3+}$) emission wavelengths - c); emission spectra ($\lambda_{exc}$ = 808 nm) - d) and luminescence decay profiles of Yb$^{3+}$ ions ($\lambda_{em}$=1060 nm) – e) of LiYO$_2$:Nd$^{3+}$,10%Yb$^{3+}$ for different

concentration of $Nd^{3+}$ ions measured at 83 K; the influence of $Nd^{3+}$ ions concentration on the $\tau_{avr}$ of $^2F_{5/2}$ state of $Yb^{3+}$ ions measured at 83 K -f).

To investigate the influence of temperature on the spectroscopic properties of $LiYO_2:Nd^{3+},Yb^{3+}$ emission spectra were recorded as a function of temperature for various $Nd^{3+}$ ion concentrations. The normalized thermal luminescence maps clearly reveal distinct shifts in the spectral positions of emission lines corresponding to electronic transitions between the Stark components of the $^2F_{5/2}$ and $^2F_{7/2}$ energy levels of $Yb^{3+}$ ions. A well-defined threshold, above which these changes become apparent for $LiYO_2:10\%Yb^{3+},1\%Nd^{3+}$, occurs at approximately 290 K, and this temperature increases with higher $Nd^{3+}$ ion concentration. This behavior, commonly observed in phase-transition-based luminescent thermometers, arises from a shift in the phase transition temperature induced by differences in the ionic radii between the dopant and host ions. An increase in temperature also affects the luminescence intensity of $Yb^{3+}$ ions (Figure 3a-e, see also Figure S1-4). Beyond a certain threshold temperature, the emission intensity gradually decreases with increasing temperature. A comparison of this behavior for different $Nd^{3+}$ ion concentrations shows that higher Nd content leads to faster thermal quenching of $Yb^{3+}$ luminescence. This phenomenon is attributed to the thermally activated nonradiative depopulation of the excited $^2F_{5/2}$ level of $Yb^{3+}$ ions through phonon-assisted $Yb^{3+}$ → $Nd^{3+}$ energy transfer. Increasing the $Nd^{3+}$ ions concentration reduces the average interionic distance, thereby enhancing the probability of this transfer. Thermal variations in the $Nd^{3+}$ ion emission are presented in Figure S5. The distinct difference in the thermal response rates of these two emission signals is particularly well illustrated by analyzing their intensity ratio:

$$LIR_1 = \frac{\int_{1068nm}^{1085nm} \left(^4F_{3/2} \rightarrow {}^4I_{11/2}\right)[Nd^{3+}]d\lambda}{\int_{940nm}^{1060nm} \left(^2F_{5/2} \rightarrow {}^2F_{7/2}\right)[Yb^{3+}]d\lambda} \quad (3)$$

It should be noted that analysis of the $Nd^{3+}$ ions emission intensity is challenging due to its relatively low signal strength and partial spectral overlap of the $^4F_{3/2} \rightarrow {}^4I_{11/2}$ band with the $^2F_{5/2} \rightarrow {}^2F_{7/2}$ transition of $Yb^{3+}$ ions. For $LiYO_2$:10%$Yb^{3+}$,1%$Nd^{3+}$ minor variations in $LIR_1$ are observed up to approximately 280 K, which corresponds to the phase transition temperature. Above this point, a slight increase in $LIR_1$ occurs. However, at higher $Nd^{3+}$ ions concentrations, this behavior changes; $LIR_1$ increases monotonically across the entire temperature range, and the rate of this increase becomes more pronounced with increasing $Nd^{3+}$ ions content. This result, associated with phonon-assisted energy transfer between interacting ions, indicates that this mechanism starts to dominate over the effects of the structural phase transition even in $LiYO_2$:10%$Yb^{3+}$,5%$Nd^{3+}$. The overall $Yb^{3+}$ luminescence intensity, however, prevents a precise assessment of the phase transition's influence on its spectroscopic behavior. Therefore, an additional analysis was conducted based on the ratio of luminescence intensities of the emission lines corresponding to the low-temperature and high-temperature regions of $LiYO_2$:$Yb^{3+}$,$Nd^{3+}$ denoted as $LIR_2$:

$$LIR_2 = \frac{\int_{967nm}^{974nm} (^2F_{5/2} \rightarrow {}^2F_{7/2}) d\lambda \left[Yb^{3+}\right]}{\int_{966nm}^{1000nm} (^2F_{5/2} \rightarrow {}^2F_{7/2}) d\lambda \left[Yb^{3+}\right]} \qquad (4)$$

The thermal dependence of $LIR_2$ exhibits far more pronounced changes compared to those observed for $LIR_1$. $LIR_2$ increases monotonically across the entire temperature range, with the steepest variation occurring near the phase transition temperature. Moreover, an increase in $Nd^{3+}$ ions concentration shifts the phase transition temperature to higher values. To quantitatively evaluate this behavior, the relative sensitivity ($S_R$) was calculated:

$$S_R = \frac{1}{LIR} \frac{\Delta LIR}{\Delta T} \cdot 100\% \qquad (5)$$

The relationships presented in Figure 3h correspond to the $LIR_2$ variations shown in Figure 3g. A sharp increase in $S_R$ is observed within the temperature range associated with the phase transition. Notably, increasing the Nd$^{3+}$ ions concentration leads to a systematic decrease in the maximum $S_R$ value from 3.2% K$^{-1}$ for 1% Nd$^{3+}$ to 0.35% K$^{-1}$ for 20% Nd$^{3+}$. Simultaneously, the temperature at which $S_R$ reaches its maximum increases linearly with Nd concentration, from 290 K (1% Nd$^{3+}$) to approximately 460 K (20% Nd$^{3+}$). These results demonstrate that, although the phase transition temperature in this system can be effectively tuned by adjusting the Nd$^{3+}$ ions content, this tuning is accompanied by a reduction in relative sensitivity.

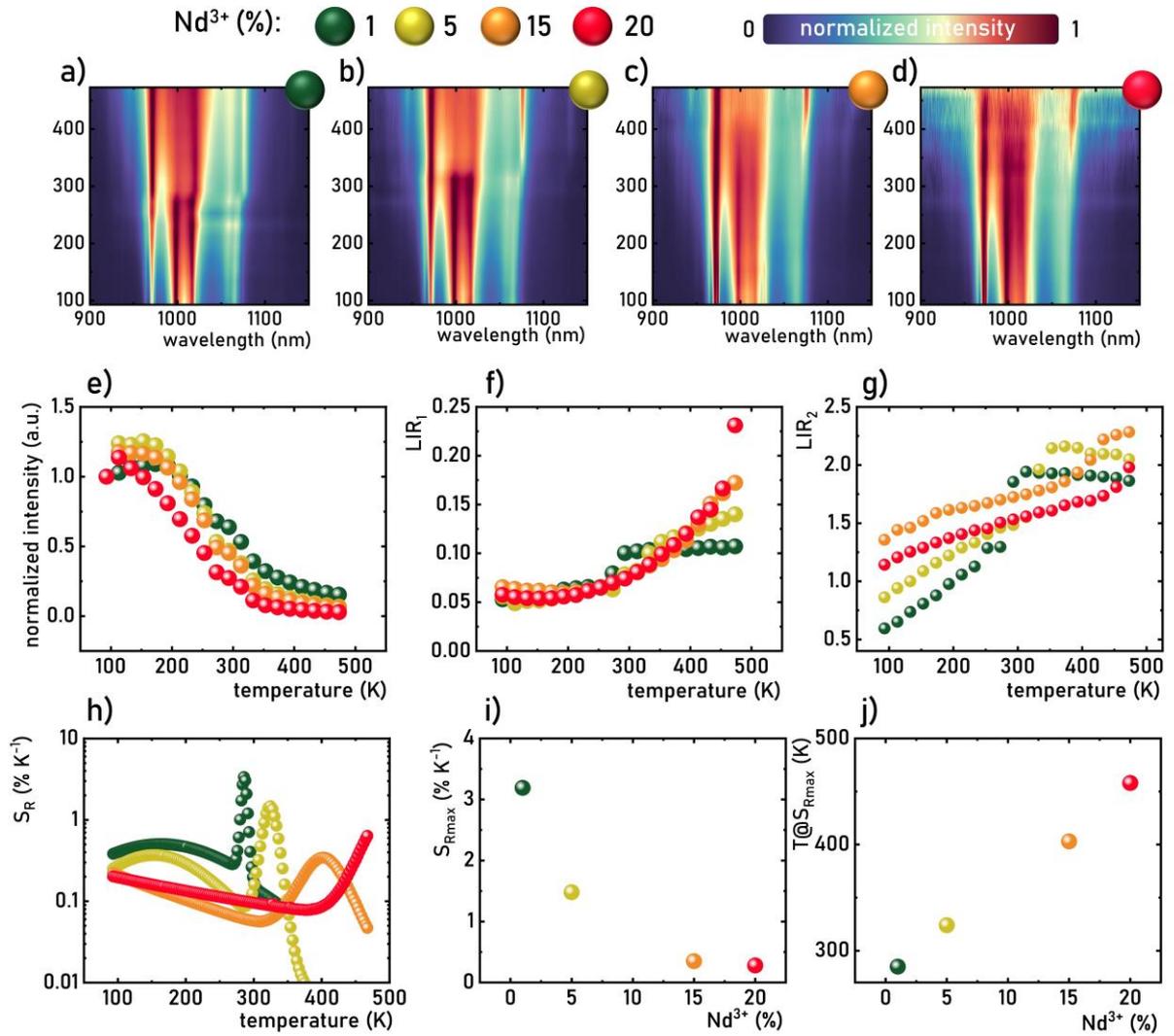

**Figure 3**. Luminescence thermal maps of normalized emission spectra of LiYO$_2$:Nd$^{3+}$,10%Yb$^{3+}$ with 1%Nd$^{3+}$ - a), 5%Nd$^{3+}$ - b), 15%Nd$^{3+}$ - c), 20%Nd$^{3+}$ - d); thermal dependence of normalized emission intensity of Yb$^{3+}$ ions – e), $LIR_1$ – f); and $LIR_2$ -g) and corresponding $S_R$ -h); the influence of Nd$^{3+}$ ions concentration on the $S_{Rmax}$ – i) and $T@S_{Rmax}$ – j).

Thermally activated energy transfer between $Yb^{3+}$ and $Nd^{3+}$ ions should be clearly reflected in the luminescence kinetics of $Yb^{3+}$ ions. The primary mechanism responsible for the thermal depopulation of excited states in lanthanide ions is multiphonon relaxation, which has relatively low probability in the case of $Yb^{3+}$ due to the large energy separation of approximately 10,000 cm$^{-1}$ between its excited $^2F_{5/2}$ level and ground $^2F_{7/2}$ level[56,57]. As a result, the luminescence kinetics of $Yb^{3+}$ are typically expected to remain unaffected by temperature variations[56,59]. However, in the case of LiYO$_2$:Yb$^{3+}$, the structural phase transition from the low-temperature to the high-temperature phase leads to a marked increase in the $\tau_{avr}$ at 300 K, with the exact temperature depending on dopant concentration[15]. As shown in Figure 4a (see also Figure S6-8), in systems co-doped with $Yb^{3+}$ and $Nd^{3+}$ ions, increasing temperature results in a progressive shortening of the luminescence decay profile. This behavior is attributed to phonon-assisted energy transfer between $Yb^{3+}$ and $Nd^{3+}$ ions, the probability of which increases with temperature. Given that the decay profiles deviate from monoexponential behavior at elevated temperatures, the $\tau_{avr}$ of the $^2F_{5/2}$ state was calculated using the procedure described in the Experimental section. The thermal evolution of $\tau_{avr}$ (Figure 4b) shows that at low $Nd^{3+}$ ions concentrations, $\tau_{avr}$ initially prolongates from 0.51 ms at 77 K to 0.6 ms at 200 K, followed by shortening to approximately 0.3 ms at 593 K. This initial prolongation of $\tau_{avr}$ may result from energy migration among excited $Yb^{3+}$ ions, and the temperature dependence of this effect suggests it is thermally activated[60–62]. The relatively low temperature at which this behavior occurs implies that energy migration may involve thermally populated higher Stark components of the $^2F_{5/2}$ state. At temperatures above 200 K, phonon-assisted energy transfer to $Nd^{3+}$ ions becomes the dominant process, leading to faster luminescence decay. As the $Nd^{3+}$ ions concentration increases, the initial $\tau_{avr}$ elongation becomes less pronounced and is no longer observed for concentrations of 15% $Nd^{3+}$ and 20%

Nd$^{3+}$, confirming that the process is related to energy migration among Yb$^{3+}$ ions. Higher Nd$^{3+}$ ions concentrations reduce the average Yb$^{3+}$ to Nd$^{3+}$ distance, thereby enhancing the efficiency of energy transfer between them. When Nd$^{3+}$ ions concentration exceeds that of Yb$^{3+}$, the Yb$^{3+}$ → Nd$^{3+}$ energy transfer overtakes energy migration as the dominant mechanism. This effect is clearly visible in the thermal dependence of the $\tau_{avr}/\tau_{avr83\,K}$ ratio (Figure 4c), which demonstrates that increasing Nd$^{3+}$ ions concentration both accelerates the rate of $\tau_{avr}$ thermal shortening and lowers the activation temperature of this process. Importantly, the structural phase transition in LiYO$_2$ is not significantly reflected in the luminescence kinetics, suggesting that the interionic energy transfer mechanism dominates the observed behavior.

Thermal variations in $\tau_{avr}$ of the Yb$^{3+}$ can be used to develop a luminescent thermometer based on luminescence kinetics. For this purpose, the relative sensitivity ($S_R$) was determined as follows:

$$S_R = \frac{1}{\tau_{avr}} \frac{\Delta \tau_{avr}}{\Delta T} \cdot 100\% \qquad (6)$$

Since reliable temperature sensing requires a monotonic response of the thermometric parameter across the operating range, $S_R$ for LiYO$_2$:Nd$^{3+}$,Yb$^{3+}$ was calculated only over the temperature interval where $\tau_{avr}$ showed consistent shortening (Figure 4d). The results indicate a distinct maximum in $S_R$, with the value depending on Nd$^{3+}$ ions concentration (Figure 4d). As the Nd$^{3+}$ ions concentration increases, the maximum $S_R$ value increases monotonically, from 0.4 % K$^{-1}$ for 5 % Nd$^{3+}$ to 1.49% K$^{-1}$ for 20% Nd$^{3+}$ (Figure 4e). This trend confirms enhanced efficiency of Yb$^{3+}$ → Nd$^{3+}$ energy transfer as the interionic distance decreases with higher doping levels. Furthermore, higher Nd$^{3+}$ ions concentrations also lower the temperature at which $S_R$ reaches its maximum ($T@S_{Rmax}$), shifting from 379 K at 5% Nd$^{3+}$ to 281 K at 20% Nd$^{3+}$ (Figure 4f). These results illustrate that thermometric performance of this luminescent thermometer can be tuned by adjusting the $Nd^{3+}$ ions concentration, offering a flexible approach to optimize luminescence thermometers based on decay kinetics.

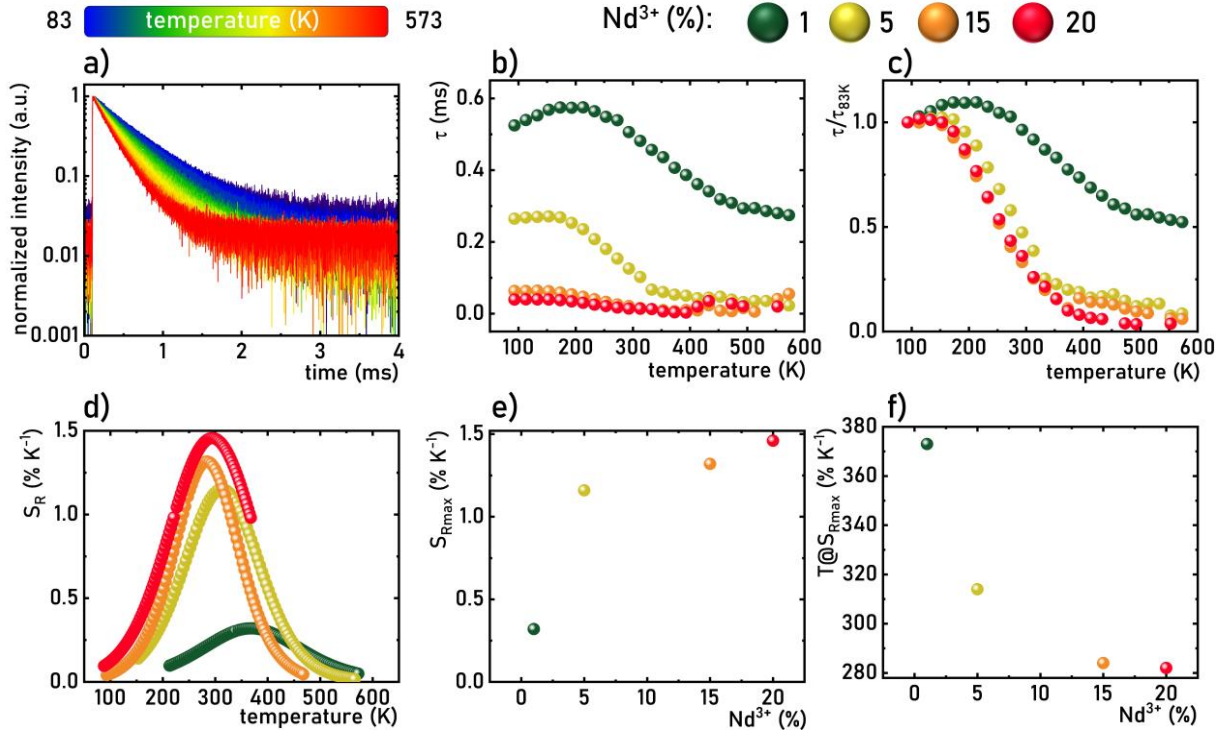

**Figure 4**. Luminescence decay profiles of $LiYO_2:1\%Nd^{3+},10\%Yb^{3+}$ ($\lambda_{exc}$=808 nm, $\lambda_{em}$=972 nm) measured as a function of temperature – a); thermal dependence of $\tau_{avr}$ – b) and $\tau_{avr}/\tau_{avr83\,K}$ – c) of $^2F_{5/2}$ state of $Yb^{3+}$ ions for different $Nd^{3+}$ ions concentration; corresponding thermal dependence of $S_R$ for lifetime-based luminescence thermometer on $LiYO_2:Nd^{3+},10\%Yb^{3+}$ - d); the influence of $Nd^{3+}$ ions concentration on the $S_{Rmax}$ - e) and $T@S_{Rmax}$ - f).

Although luminescence kinetics is considered one of the most reliable thermometric parameters due to its low sensitivity to variations in the optical properties of the medium containing the phosphor, its use in thermal imaging presents certain limitations. The primary challenge lies in the fact that measuring luminescence decay profiles requires acquisition at each individual point within the observed area, which is highly time-consuming and may become impractical in dynamic conditions where temperature changes rapidly. Under such circumstances, obtaining reliable data is significantly hindered. As an effective alternative to full luminescence decay profile analysis, a recently introduced strategy employs the ratio of luminescence intensities collected within two distinct time gates[63,64] (Figure 5a). In this approach, thermal shortening of the decay profile leads to differences in the temperature dependence of the signal intensity recorded in each gate. As demonstrated in previous studies, both the duration and temporal separation of the gates can influence the resulting temperature sensitivity[63]. To evaluate the applicability of this method for the LiYO$_2$:Nd$^{3+}$,Yb$^{3+}$ system, two time gates were arbitrarily defined: *G1* (0 to 0.25 ms) and *G2* (0.25 to 0.5 ms). The analysis revealed that, for 1% Nd$^{3+}$ doping, the signal recorded in *G1* exhibits low sensitivity to temperature changes (Figure 5b). This is attributed to the relatively long $\tau_{avr}$ observed at this dopant level, such that *G1* captures the time window where the luminescence intensity remains high and relatively stable. As the Nd$^{3+}$ ions concentration increases, a gradual thermal decrease in *G1* signal intensity is observed, consistent with the behavior noted in the $\tau_{avr}$ -based analysis (Figure 4). In contrast, the signal recorded in *G2* shows higher thermal variability for low Nd$^{3+}$ ions concentrations (Figure 5c). However, for samples co-doped with 15 % and 20 % Nd$^{3+}$, only slight thermal dependence is observed in *G2*, primarily due to the fact that short $\tau_{avr}$ values result in minimal emission intensity within the *G2* time window. The markedly different thermal behavior of the *G1* and *G2* signals enables the use of their ratio as a ratiometric parameter, defined as *LIR$_3$*:

$$LIR_3 = \frac{\int\limits_{0ms}^{0.25ms} \left(^2F_{5/2} \to {}^2F_{7/2}\right)[Yb^{3+}]dt}{\int\limits_{0.25ms}^{0.5ms} \left(^2F_{5/2} \to {}^2F_{7/2}\right)[Yb^{3+}]dt} \qquad (7)$$

The results presented in Figure 5d show that for low Nd$^{3+}$ ions concentrations, *LIR$_3$* exhibits modest changes with temperature, with a maximum variation around 400 K. As the Nd$^{3+}$ ions concentration increases, the thermal response of *LIR$_3$* becomes more pronounced and dynamic, which is reflected in the corresponding $S_R$, shown in Figure 5e. As in the case of $\tau_{avr}$-based thermometry, $Nd^{3+}$ ions concentration acts as a tuning parameter for thermometric performance. The maximum $S_R$ value increases with $Nd^{3+}$ ions content, rising from 0.35 % K$^{-1}$ for 1% $Nd^{3+}$ to 2.1 % K$^{-1}$ for 15% $Nd^{3+}$ (Figure 5f). A further increase in dopant concentration to 20% $Nd^{3+}$ slightly reduces $S_{Rmax}$ to 1.5% K$^{-1}$. Similar to $\tau_{avr}$-based observations, higher $Nd^{3+}$ concentrations shift the temperature corresponding to $S_{Rmax}$ toward lower values, from approximately 395 K at 1% $Nd^{3+}$ to around 310 K at 20% $Nd^{3+}$ (Figure 5g). It is worth noting that this gated detection method leads to a slight increase in the temperature at $S_{Rmax}$, by approximately 20 K, compared to values obtained via $\tau_{avr}$-based analysis. Overall, the results clearly demonstrate that thermal changes in the luminescence kinetics of $LiYO_2:Nd^{3+},Yb^{3+}$ can be effectively utilized for both temperature sensing and thermal imaging.

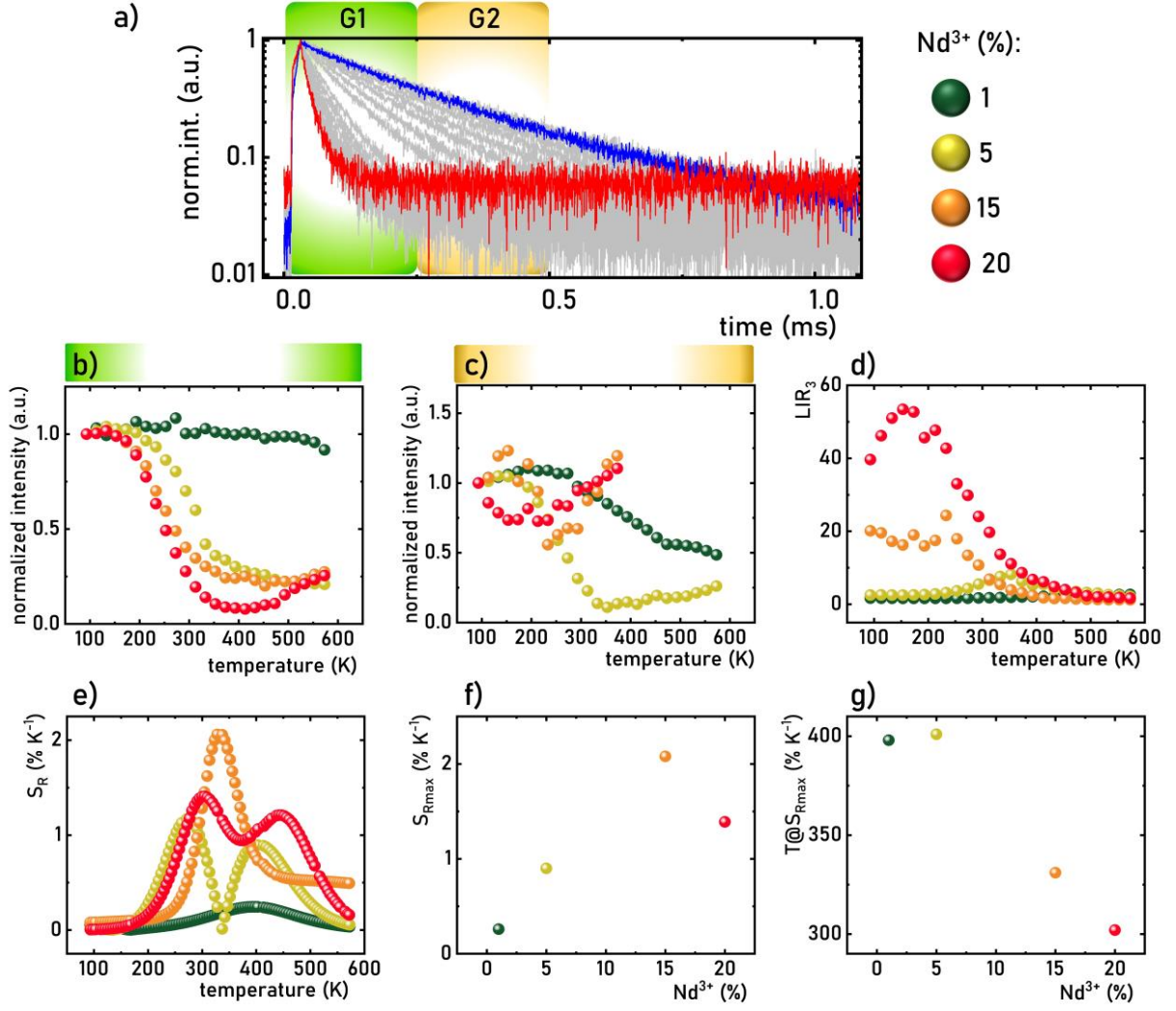

**Figure 5**. Luminescence decay profiles of LiYO$_2$:1%Nd$^{3+}$,10%Yb$^{3+}$ measured as a function of temperature with marked two time gates (G1 – 0-0.25 ms, G2 0.25 – 0.50 ms) used in the calculations – a); thermal dependence of the emission intensity signal registered in the *G1* – b) and *G2* – c) time gates; thermal dependence of *LIR$_3$* – d); and corresponding *S$_R$* – e); the influence of the Nd$^{3+}$ ions concentration on the *S$_{Rmax}$* – f) and *T@S$_{Rmax}$* – g) of the luminescence thermometer based on time gates.

## Conclusions

In this work, the spectroscopic properties of LiYO$_2$ codoped with Nd$^{3+}$ and Yb$^{3+}$ ions were investigated as a function of temperature to develop a luminescent thermometer based on both ratiometric and lifetime readout strategies. In the studied system, a thermally induced structural phase transition from a monoclinic to a tetragonal phase was identified. It was found that increasing the $Nd^{3+}$ ion concentration raises the phase transition temperature without significantly affecting the morphology of the samples. Analysis of the temperature-dependent emission spectra of $LiYO_2:Nd^{3+},Yb^{3+}$ revealed notable changes in the shape of the emission band associated with the $^2F_{5/2} \rightarrow {}^2F_{7/2}$ electronic transition, coinciding with the structural phase transition. These changes are manifested by shifts in the spectral positions of emission lines corresponding to transitions between the Stark components of the $^2F_{5/2}$ and $^2F_{7/2}$ energy levels of $Yb^{3+}$ ions. The thermally induced modifications in the emission band shape of $Yb^{3+}$ ions were used to construct a ratiometric luminescence thermometer. The relative sensitivity and the temperature at which the maximum sensitivity is achieved can be tuned by adjusting the $Nd^{3+}$ ion concentration. Specifically, the sensitivity decreases from 3.1% $K^{-1}$ for 1% $Nd^{3+}$ to 0.4% $K^{-1}$ for 20 % $Nd^{3+}$, while the corresponding temperature increases from 290 K to approximately 470 K, respectively. Additionally, rising temperature activates phonon-assisted energy transfer from $Yb^{3+}$ to $Nd^{3+}$ ions. The efficiency of this transfer increases with decreasing average interionic distance, which is achieved by increasing the $Nd^{3+}$ ions concentration. As a result, the $\tau_{avr}$ of $Yb^{3+}$ shortens with temperature, and this process becomes the dominant mechanism, surpassing the effects of the structural phase transition. With increasing $Nd^{3+}$ ions content, the rate of thermal shortening of the $\tau_{avr}$ increases, while the temperature range over which this effect is observed becomes narrower. This leads to an increase in $S_R$ of lifetime-based luminescent thermometer from 0.3% $K^{-1}$ for 1% $Nd^{3+}$ to 1.5% $K^{-1}$ for 20% $Nd^{3+}$, and a corresponding decrease in the temperature at maximum sensitivity from 378 K to 280 K. The dominant influence of the structural phase transition on spectral shape, combined with the role of $Yb^{3+} \rightarrow Nd^{3+}$ energy transfer in the thermal evolution of luminescence kinetics, enables the development of a dual-mode sensor. This system offers two distinct readout methods with different thermal operating ranges. Moreover, the potential of $LiYO_2:Nd^{3+},Yb^{3+}$ for temperature sensing based on the ratio of luminescence intensities recorded within two time gates was explored. This time-gated approach allows not only for thermal imaging but also results in an increase in both the maximum relative sensitivity and the temperature at which it occurs compared to the conventional lifetime-based approach. Notably, a maximum sensitivity of 2.1 % K$^{-1}$ was achieved for the LiYO$_2$:10% Yb$^{3+}$, 15 % Nd$^{3+}$ at 340 K. In conclusion, the study demonstrates that LiYO$_2$ co-doped with Yb$^{3+}$ and Nd$^{3+}$ is a promising candidate for multimodal luminescent temperature sensing, offering tunable performance and applicability across a range of thermal conditions

**Acknowledgements**

This work was supported by the National Science Center (NCN) Poland under project no. DEC-UMO-2022/45/B/ST5/01629.